\documentclass[12pt,3p,preprint,sort&compress]{elsarticle}
\pdfoutput=1
\usepackage[colorlinks=true]{hyperref}

\usepackage{amsmath}
\usepackage{amssymb}
\usepackage{mathtools}
\usepackage[utf8]{inputenc}

\newcommand{\e}{{\mathrm{e}}}

\renewcommand{\l}{\left(}
\renewcommand{\r}{\right)}

\newcommand{\be}{\begin{equation}}
\newcommand{\ee}{\end{equation}}
\newcommand{\ba}{\begin{align}}
\newcommand{\ea}{\end{align}}
\newcommand{\bg}{\begin{gather}}
\newcommand{\eg}{\end{gather}}
\newcommand{\bseq}{\begin{subequations}}
\newcommand{\eseq}{\end{subequations}}

\begin{document}
\begin{flushright}
	INR-TH-2021-024
\end{flushright}

\title{Revisiting PS191 limits on sterile neutrinos} 
\author[inr,mpti]{Dmitry Gorbunov}
\ead{gorby@ms2.inr.ac.ru}
\author[inr]{Igor Krasnov}
\ead{iv.krasnov@physics.msu.ru}
\author[inr,mpti,lpnhe]{Sergey Suvorov}
\ead{suvorov@inr.ru}
\address[inr]{Institute for Nuclear Research of Russian Academy of Sciences, 117312 Moscow, Russia}
\address[mpti]{Moscow Institute of Physics and Technology, 141700 Dolgoprudny, Russia}
\address[lpnhe]{LPNHE Paris, Sorbonne Universite, Universite Paris Diderot, CNRS/IN2P3, Paris 75252, France}
\begin{abstract}
      We perform Monte Carlo simulations of the sterile neutrino signal at the fixed target experiment PS191 operated on a proton beam of 19.2\,GeV at CERN in the eighties. We find that the strongest bounds the PS191 could obtain are significantly lower than what they published, and now are obsolete being surpassed by recent T2K, NA62, E949, TRIUMF and PIENU experiments. 
\end{abstract}
\date{}

\maketitle

{\bf 1.} 
Sterile neutrinos--massive fermions singlet with respect to the Standard Model gauge group--are one of the viable models of physics beyond the Standard Model, originally suggested to explain the smallness of the active neutrino masses via the seesaw mechanism\,\cite{Minkowski:1977sc,Yanagida:1980xy}, but also capable of generating the baryon asymmetry of the Universe and playing a role of dark matter (say, within the $\nu$MSM setup\cite{Shaposhnikov:2008pf}). Phenomenology of the sterile neutrino sector is solely determined by its mass scale and mixing between the sterile and active neutrino components, see e.g.\cite{Bondarenko:2018ptm}. The latter implies that the sterile neutrinos (called also as Heavy Neutral Leptons, HNLs) can be involved in weak interactions, and, in particular, be produced in weak decays and subsequently decay weakly into the Standard Model particles, if kinematically allowed. These processes can be searched for in particle physics experiments, and many have been performed aiming at the discovery of the sterile neutrino sector.

One of such experiments, and probably the first dedicated to the searches for HNLs, is PS191\,\cite{Bernardi:1985ny,Bernardi:1986hs,Bernardi:1987ek} at CERN. It exploited the highly intensive beam of 19.2\,GeV protons incident on the target.  The produced kaons might decay into HNLs, which subsequently decay in the vacuum vessel at some distance from the target. The decays into states with a couple of charged particles were adopted as the HNL signatures. The PS191 announced no signal events, and so placed upper limits on the sterile-active mixing angles for HNL masses in 100-400\,MeV range\,\cite{Bernardi:1985ny,Bernardi:1987ek}. The experiment was simple but well designed. Though it operated in the eighties, the obtained limits remained strongest for many years and are still considered the strongest for particular mass regions. 

However, recent analysis\,\cite{Arguelles:2021dqn} of the HNL phenomenology (mixing with muon neutrinos) at hodoscopic neutrino detectors after comparing the basic characteristics (beam energy, collected number of protons on target $N_{POT}$, detector size and position, etc) of the old PS191 and presently operating T2K questioned the reliability of the former results, given the latter exhibits better characteristics, but failed to surpass the PS191 limits. While the PS191 detector is quite different from what T2K uses, and much more suitable for HNL signatures, the arguments presented in Ref.,\cite{Arguelles:2021dqn} look convincing enough to investigate the matter. Certainly, it would be impossible to trace the whole experimental procedure, but the HNL signal can be simulated with reasonable accuracy, and so {\it accepting there were no signal events in PS191}, one can estimate the corresponding limits on the sterile-active mixing angles and compare them with those published by PS191 collaborations in Refs.\,\cite{Bernardi:1985ny,Bernardi:1987ek}.   

\begin{figure}[!htb]
    \centering
    \includegraphics[width=\textwidth]{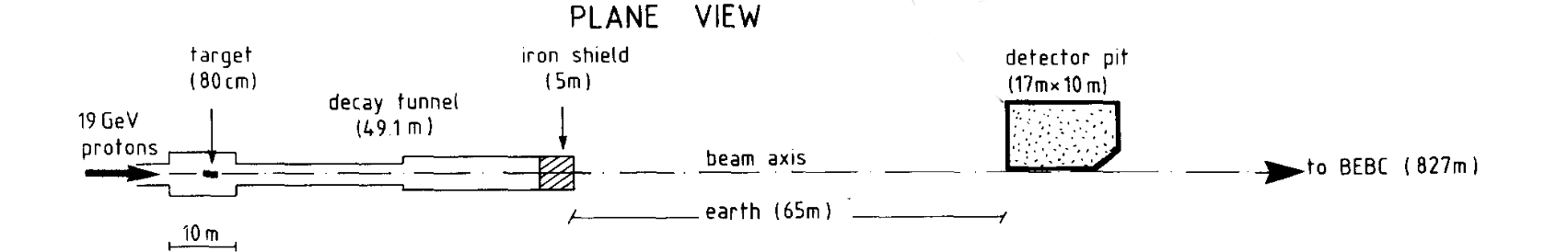}
    \caption{The layout of the PS191 experiment at CERN, adopted from Ref.\,\cite{Bernardi:1985ny}.}
    \label{fig:PS191}
\end{figure}

\begin{figure}[!htb]
    \centering
    \includegraphics[width=0.8\textwidth]{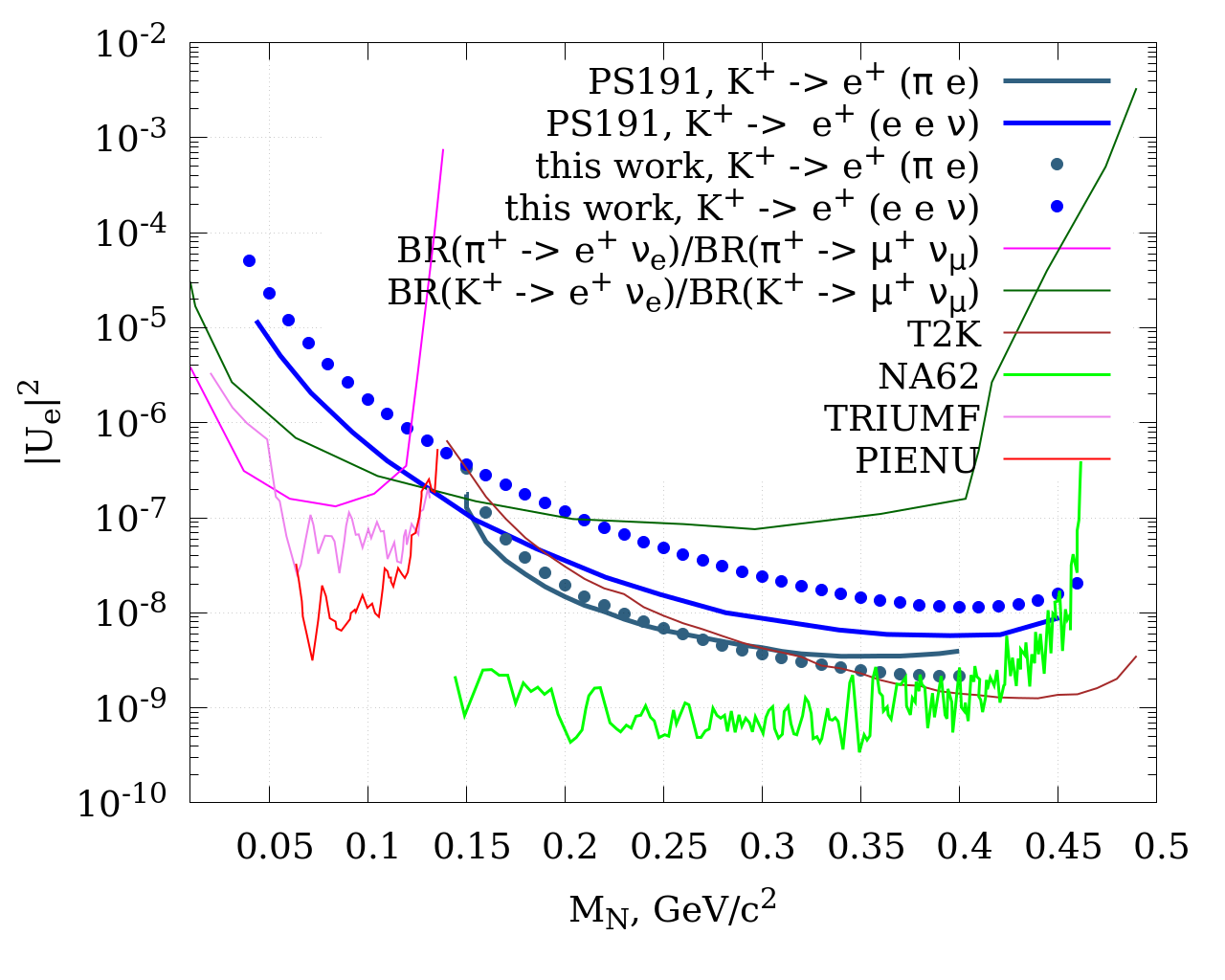}
    \caption{Limits on sterile neutrino mixing with electron neutrino obtained in this work (dots) from simulations of the HNL signal at PS191 experiment and bounds published by PS191 collaboration. They are obtained for particular processes indicated on the plot. There are also constraints on the same mixing from other experiments  (T2K\cite{T2K:2019jwa}, NA62\cite{NA62:2020mcv}, TRIUMF\cite{Britton:1992xv}, PIENU\cite{PIENU:2017wbj}) and bounds\,\cite{Bryman:2019bjg} inferred from consistence of measured ratios of pion and kaon leptonic branchings with the SM predictions.}
    \label{fig:ee}
\end{figure}

{\bf 2.} In this letter, we perform the simulation of the HNL signal at the PS191 experiment, which layout is sketched in  Fig.\,\ref{fig:PS191}. 
It utilized protons of energy 19.2\,GeV incident on a beryllium target (a cylinder 6\,mm in diameter and 80\,cm long). The proton beam was produced by PS (Proton Synchrotron) of CERN. The protons had enough energy to produce light mesons, and for the HNL searches kaons were of special interest.\footnote{There were also PS191 limits for HNL produced in pion decays, but they have been surpassed by later experiments, see Fig. \ref{fig:ee}.}   The sterile neutrino detection scheme can be summarized as follows. The kaons travel along the decay tunnel (filled with Helium), see Fig.\,\ref{fig:PS191}, with rectangular cross section of 5\,m$\times$2.8\,m at the downstream end. Some kaons might scatter off the tunnel walls or the 5\,m long iron dump closing the tunnel, where strangeness can be transferred to baryons. The survived kaons decay through weak interactions. 
Then each standard decay mode with active neutrino in the final state can be supplemented by a similar one with HNL instead of the neutrino. Two-body kaon decays yield the dominant contribution to the HNL production, the branching to charged lepton $l=e,\mu$ and sterile neutrino $N$ reads \cite{Shrock:1980ct,Gorbunov:2007ak} 
\begin{equation}
    \label{Br-2}
    \begin{split}
&\text{Br}(K^\pm \to l^\pm N)  = |U_{l}|^2 \frac{\tau_K}{8 \pi}  G_F^2 f_K^2 M_K M_N^2 |V_{us}|^2\\
&\times\l1 - \frac{M_N^2}{M_K^2} + 2 \frac{M_l^2}{M_K^2} + \frac{M_l^2}{M_N^2} \l1-\frac{M_l^2}{M_K^2}\r\r\times \sqrt{\l 1 + \frac{M_N^2}{M_K^2} - \frac{M_l^2}{M_K^2}\r^2 - 4\, \frac{M_N^2}{M_K^2}}\,.
\end{split}
\end{equation}
with $f_K=155.6$\,MeV and $\tau_K=1.238\times 10^{-8}$\,s being respectively the kaon decay constant and lifetime\,\cite{Zyla:2020zbs}, and $M_l$ denoting the charged lepton mass. Remarkably, the branching is proportional to the mass squared of the heaviest particle in the final state, either charged or neutral lepton. Note also, that HNLs can be produced in 3-body semileptonic decay of kaons, but the corresponding branching is suppressed by the phase space factor and the mode is kinematically allowed for a shorter range of HNL masses. Thus for any $M_N$ its  contribution to the HNL signal is much smaller than that of 2-body leptonic mode and we neglect it in accordance with PS191 analysis\,\cite{Bernardi:1985ny,Bernardi:1987ek}. The two-body mode is enhanced with respect to the SM case, since its amplitude is proportional to the mass of the heaviest fermion, that is the HNL mass in our case. 

Then the HNL propagates without any noticeable attenuation and decays weakly (due to active-sterile mixing). If the decay happened inside the detector it could be observed. The detector length and cross sections were $\Delta l=12$\,m and 6\,m$\times$3\,m, respectively. It was placed at $2.29^\circ$ offset from the proton beam line and distance of $d=128$\,m from the target. Since the detector was designed to recognize the tracks of charged particles, only two- and three-body final states of the HNL decays were accessible. The hit in the electromagnetic calorimeter (equipped with hodoscope) installed downstream the detector was utilized as a trigger, and the associated pair of charged tracks coming from a single point provides the signal signature. For the HNL lighter than kaon the relevant decay modes are two-body into charged pion and lepton (electron or muon) and three-body into an active neutrino and a couple of charged leptons, electrons or muons, $l=e\,\mu$. Actually, only $\pi^\pm e^\mp$, $\e^+e^-$ and $\e^\pm\mu^\mp$ pairs were searched for in the original paper, and hence here we account only for them as well. The corresponding decay rates are given by\,\cite{Gorbunov:2007ak} (we neglect all small terms proportional to the electron mass)  
\begin{align}
\label{2-body-mu}
\Gamma\l N\to \pi^{+} \mu^{-} \r&=\frac{|U_\mu|^2}{16\pi}G_F^2|V_{ud}|^2f_\pi^2
M_N^3\cdot\l \l 1-\frac{M_\mu^2}{M_N^2}\r^2-\frac{M_\pi^2}{M_N^2}
\l 1+ \frac{M_\mu^2}{M_N^2}\r \r\\&\times 
\sqrt{\l 1-\frac{\l M_\pi -M_\mu\r^2}{M_N^2} 
\r \l 1-\frac{\l M_\pi +M_\mu\r^2}{M_N^2} \r}\;,\\
\label{2-body-e}
\Gamma\l N\to \pi^{+} e^{-} \r&=\frac{|U_e|^2}{16\pi}G_F^2|V_{ud}|^2f_\pi^2
M_N^3\cdot\l 1-\frac{M_\pi^2}{M_N^2}\r^2\;,\\
\label{3-body-eee}
\Gamma\l N\to \nu_e e^+e^- \r&=
\frac{G_F^2M_N^5}{192\pi^3}\cdot |U_e|^2 \cdot   \frac{1}{4}\l 1+4\sin^2\theta_w+8\sin^4\theta_w\r\;,
\\
\label{3-body-muee}
\Gamma\l N\to \nu_\mu e^+e^- \r&=
\frac{G_F^2M_N^5}{192\pi^3}\cdot |U_\mu|^2 \cdot 
  \frac{1}{4}\l 1-4\sin^2\theta_w+8\sin^4\theta_w\r\;,
\\
\label{3-body-muemu}
\Gamma\l N\to e^-\mu^+  \nu_\mu \r&=
\frac{G_F^2M_N^5}{192\pi^3}\cdot |U_e|^2 \l
1-8\frac{M^2_\mu}{M^2_N}+8\frac{M_\mu^6}{M_N^6}-\frac{M_\mu^8}{M_N^8} -12\frac{M_\mu^4}{M_N^4} \log \frac{M_\mu^2}{M_N^2} \r\;,\\
\label{3-body-emue}
\Gamma\l N\to \mu^-e^+  \nu_e \r&=
\frac{G_F^2M_N^5}{192\pi^3}\cdot |U_\mu|^2 \l
1-8\frac{M^2_\mu}{M^2_N}+8\frac{M_\mu^6}{M_N^6}-\frac{M_\mu^8}{M_N^8} -12\frac{M_\mu^4}{M_N^4} \log \frac{M_\mu^2}{M_N^2} \r\;,
\end{align}
 where $G_F$ is Fermi coupling constant, $V_{ud}=0.9737$ is the CKM matrix element, $\sin^2\theta_w=0.2313$ and pion decay constant $f_\pi=130.41$MeV\,\cite{Zyla:2020zbs}. Here, following the PS191 analyses,  we suppose that HNL are Dirac fermions (for the Majorana case the charged conjugated decay modes must be added). Recall that PS191 considered only processes initiated by charged currents (exchanges of virtual $W$-boson) and fully neglected contributions of neutral currents (exchanges of virtual $Z$-boson). Hence, they did not account the decay mode $N\to \nu_\mu e^+e^-$ (and similar one to tau-neutrino) and used for the decay rate of $N\to \nu_e e^+e^-$ another formula, where only charged current contribution was accounted. It can be obtained from the formula above for decay rate into $N\to e^-\mu^+\nu_\mu$ (initiated purely by the charged current interactions) by setting $M_\mu=0$. Numerically the deviation is 
 \begin{equation}
 \label{corr-CC-NC}
\eta_{NC}= \frac{1}{4}\l 1+4\sin^2\theta_w+8\sin^4\theta_w\r\approx 0.59
 \end{equation}
 and we correct the PS191 bounds for this factor in due course. 
 
{\bf 3.} 
The experiment PS191 had observed no signal events. The published limits\,\cite{Bernardi:1985ny,Bernardi:1987ek} were based on the total statistics of $N_{\text{POT}}=0.86\times10^{19}$. 
Roughly two-third of the total statistics, namely $N_{\text{POT}}^{\text{on}}=0.56\times10^{19}$, had been collected with a focusing magnets switched on. These magnets had been installed downstream of the target and were designed to focus the positively charged mesons on their way along the decay tunnel. We failed to find enough information in the literature about this magnet system enough to simulate the PS191 operation in the mode with the magnetic field switched on (we call this mode ``on''). Instead, we assumed the ultimate focusing system, which in the ``on''-mode concentrates all the charged kaons right on the beam-line. Then they propagate along the tunnel and decay. This operation can be easily simulated, and yet the HNL signal can be only amplified in this way. Aiming at estimating the strongest limit PS191 could achieve we find this simplification viable and adopt it in our simulations of the HNL signal in the ``on''-mode. The contribution of the negatively charged kaons in this mode we can calculate fully neglecting the magnetic field. Since it is supposed to ``defocus'' $K^-$, our treatment can only amplify the HNL signal. The rest of statistics $N_{\text{POT}}^{\text{off}}=0.30\times10^{19}$, were collected with magnetic field switched off, and the simulations of PS191 operation in this mode is rather straightforward. We find this strategy for performing simulations reasonably simple yet reliable and conservative in the sense, that we expect to obtain the limits definitely not weaker than the true limits PS191 could obtain. 

\begin{figure}[!htb]
    \centering
    \includegraphics[width=0.7\linewidth]{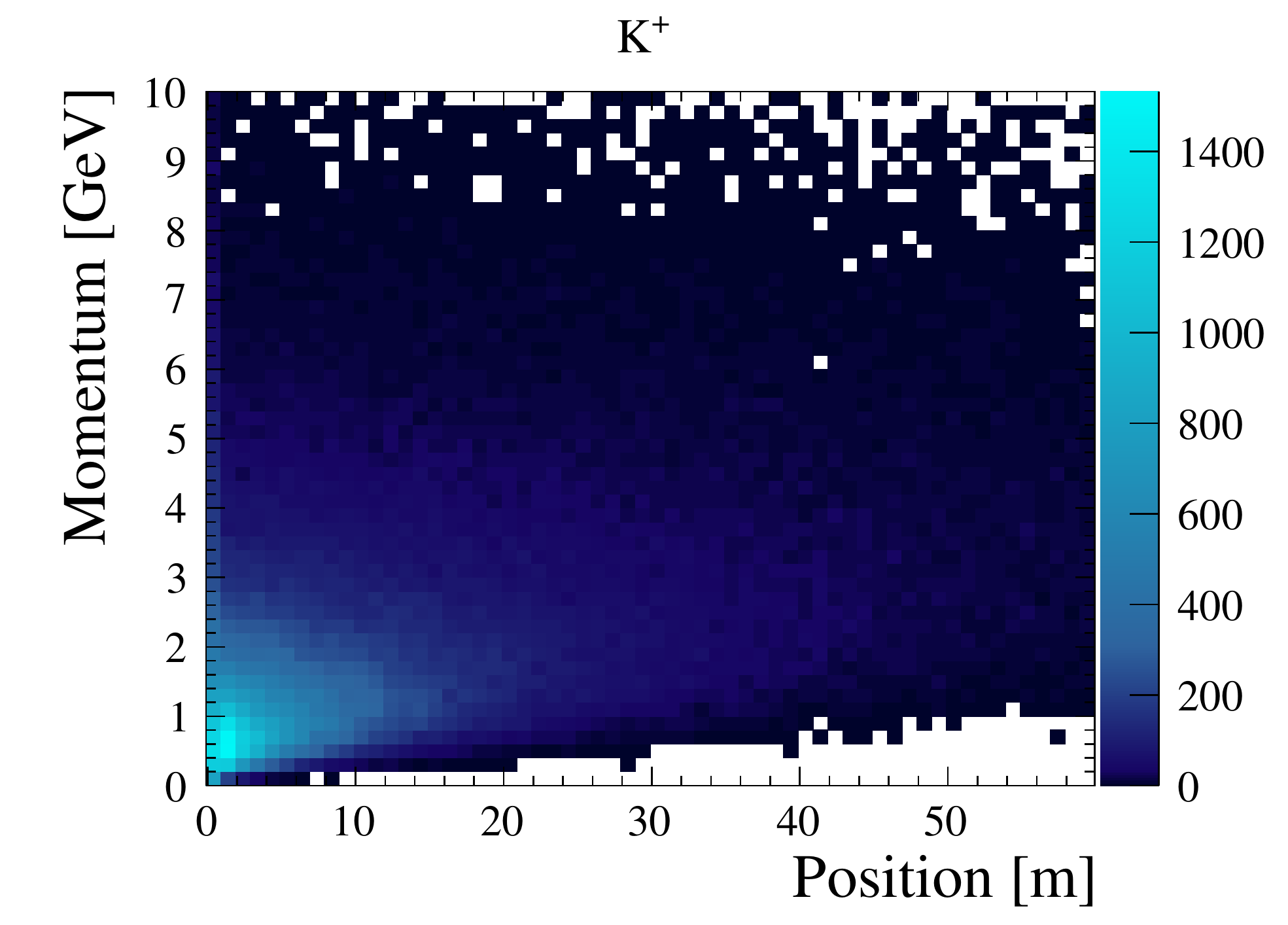}
    \caption{Distributions of 3-momentum at the decay point and travel distance of the positively charged kaons as we adopted for the ``on''-mode.}
    \label{fig:kaon_mom_dist}
\end{figure}

We estimate the expected number of HNL signal events in the PS191 detector by making use of the Monte Carlo simulations of the kaon production by $N_{sim}=2\times 10^6$ protons of energy 19.2\,GeV, which we have  performed in Ref.\,\cite{Gorbunov:2021ccu} for investigation of PS191 sensitivity to hypothetical light scalars. The kaon production, further propagation and scattering in the target, decay tunnel (walls) and iron dump till the weak decay have been simulated with GEANT4 toolkit\,\cite{Agostinelli:2002hh} with BERTini and QGSP models utilized for low and high energy strong interactions respectively. Many details (numbers, particle distributions over momenta and decay points, etc) of the simulations can be found in Ref.\,\cite{Gorbunov:2021ccu}. We accounted for the secondary kaon production in the tunnel walls, iron dump and soil (we approximated it as sand). Say, in the first interaction the incident proton produces about 0.065 $K^+$ and 0.032 $K^-$, see Tab.\,1 in \cite{Gorbunov:2021ccu} for the total kaon budget. 
We use those simulated $K^-$ for both ``on''- and ``off''-modes as we argued above. The simulated $K^+$ we use for the ``off''-mode only. For the ``on''-mode we performed another simulation of $N_{\text{sim}}=2\times 10^6$ protons with the same QCD models but assuming another geometry: an empty space downstream the target. Then all the positively charged kaons (we have about $137\times 10^3$ in the simulation) propagate outside the target rectilinearly and finally decay. We take all these trajectories and rotate them to put along the beam axis (assuming, as we argued above, that it would be the job of an ideal focusing system, which most probably conserves the particle energy). Most trajectories end in the decay tunnel: less than ten percent of them  reach the position of iron dump, and we treat them as decaying there like in a vacuum, i.e. without any scattering. This can certainly only increases the signal. The 3-momenta of $K^+$ at the moment of decay as well as kaon travel distances from the target are shown in Fig.\,\ref{fig:kaon_mom_dist}. 
The kinematics is shared between ``on'' and ``off'' modes, but in the case of the first one, the direction of the kaons was modified as described above. In the ``off''-mode the charged kaon distributions over production point, decay point and 3-momenta at the decay point can be found in Figs.\,3-5 of Ref.\,\cite{Gorbunov:2021ccu} for the same number of simulated protons.

\begin{figure}[!htb]
    \centering
    \includegraphics[width=0.8\textwidth]{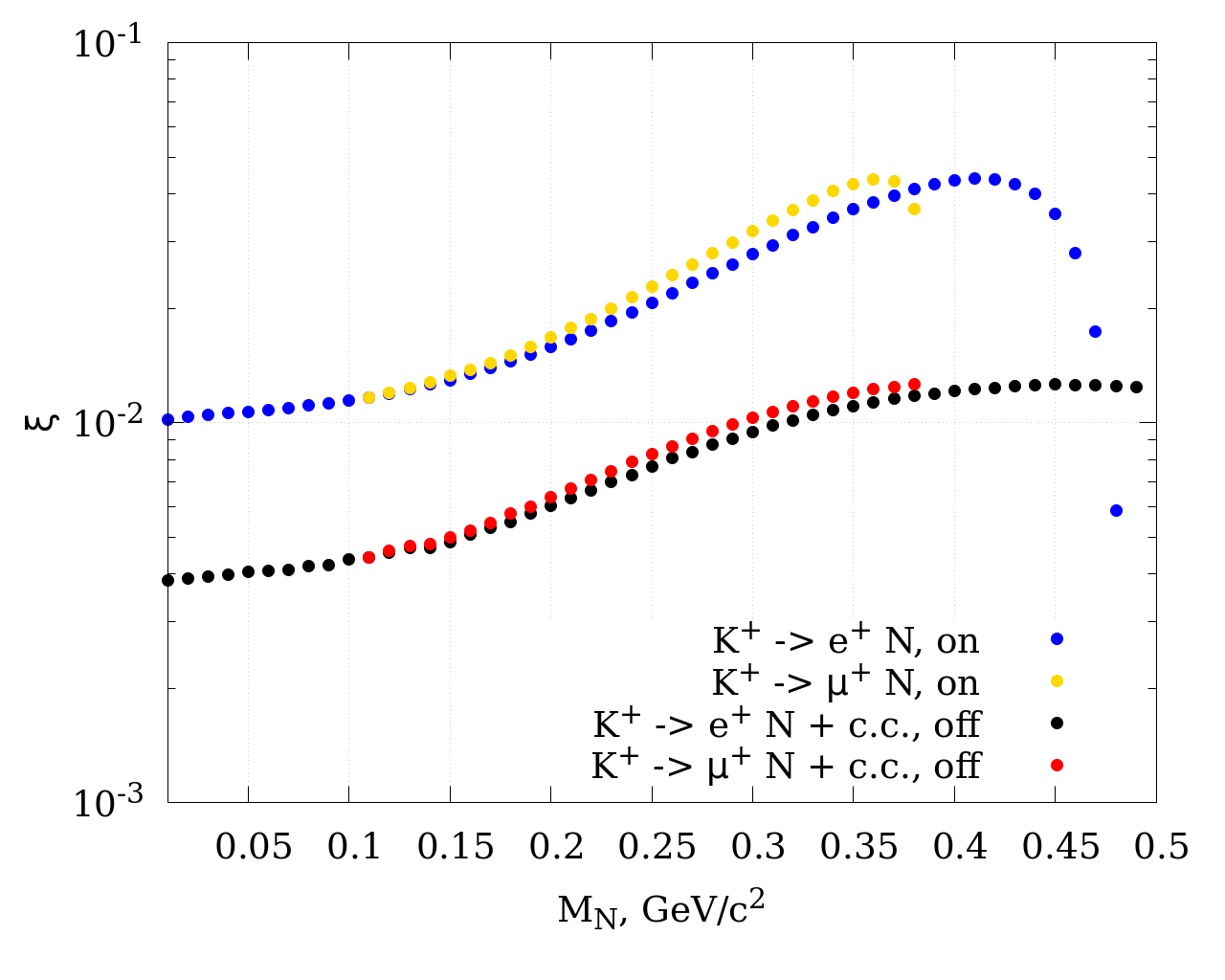}
\caption{The geometrical factor $\xi$ for the HNLs as a function of its mass for HNLs produced in charged kaon decays. ``On''-mode stands for our simulation of working focusing system. This is the fraction of all HNL which trajectories pass through the PS191 detector.}
    \label{fig:gfactor}
\end{figure}

In both simulations (designed for ``on''- and ``off''-modes) we treat the kaon decay as that to HNL and charged lepton with branching\,\eqref{Br-2} and the HNL 3-momentum uniformly distributed over the 2-sphere in the kaon rest frame. Then we perform the Lorentz boost to the laboratory frame and check whether the trajectory passes through the PS191 detector (see Fig.\,\ref{fig:gfactor} 
for the corresponding geometrical factor). If so, we estimate the probability of HNL to decay inside the detector considering each of the decay modes of eqs.\,\eqref{2-body-mu}-\eqref{3-body-emue} separately.  Following the original requirements from PS191 analysis we consider the efficiency to be 100$\%$ under condition that both charged particles produced in HNL decay pass through the hodoscope. Each of the kaon from simulation and each of the HNL we can use several times rescaling the number of events in the detector since both the HNL production rates and HNL decay rates are proportional to the corresponding squared mixing. Finally, the number of the signal events in the PS191 detector is given after rescaling of the number of simulated protons $N_{sim}$ to the numbers of protons in each mode, $N^\text{on}_\text{POT}$ and $N^\text{off}_\text{POT}$.    We checked our procedure by taking the limit of massless HNL and reproducing the muon neutrino spectrum at the detector obtained with the GEANT4 simulations.

{\bf 4.} No signal events had been observed in the PS191 experiment, and so for each mode and each HNL mass, we find the mixing angle at which the number of simulated events equals 2.3: according to the Poisson statistics larger mixing angles are excluded at the 90\%\,CL limit. In the original papers, the PS191 collaboration placed the exclusion limits at the same confidence level, 90\%\,CL. 

The results for $|U_e|^2$, $|U_\mu|^2$ and $|U_eU_\mu|$ are presented in Figs.\,\ref{fig:ee}, 
\begin{figure}[!htb]
    \centering
    \includegraphics[width=0.8\textwidth]{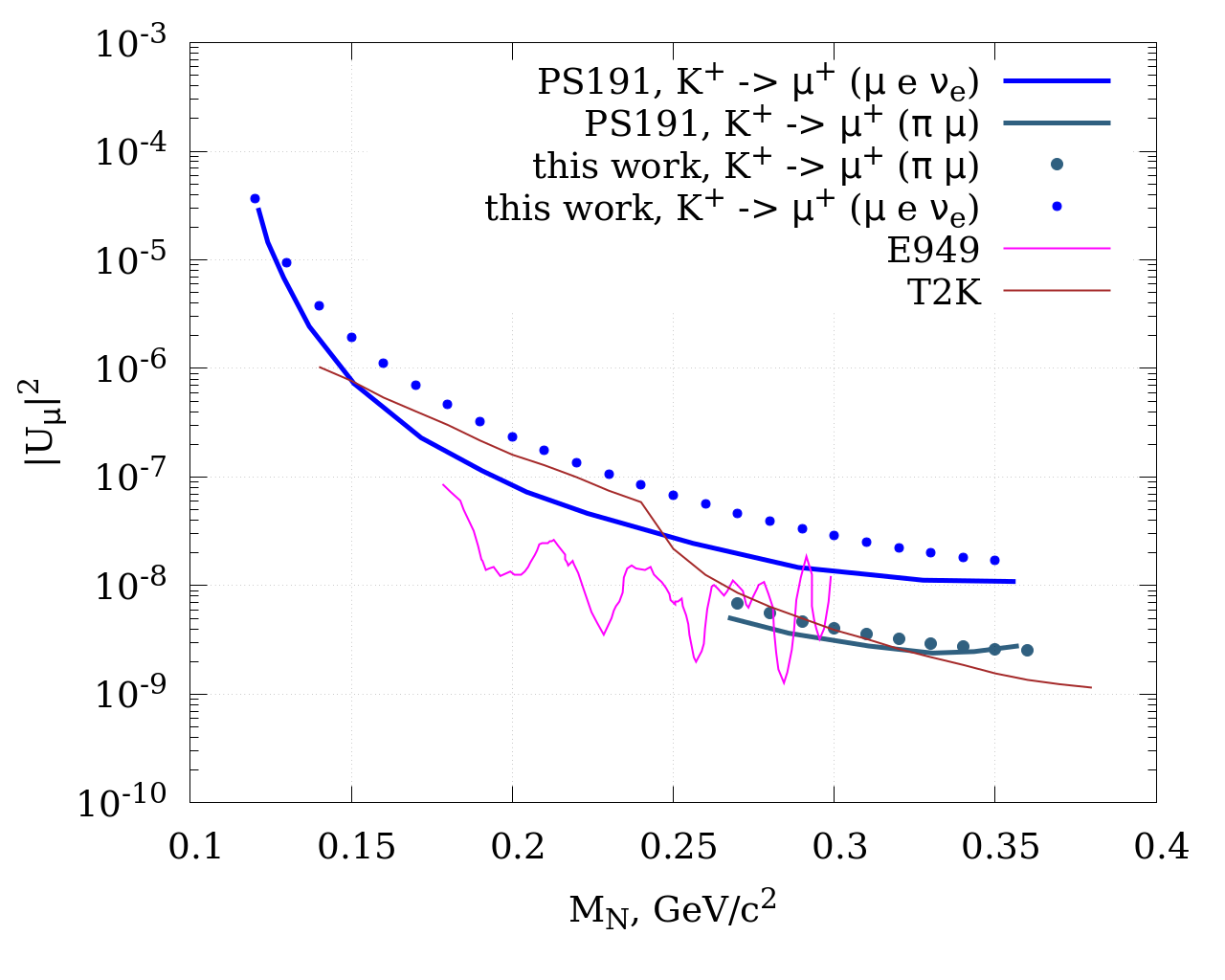}
    \caption{Limits on sterile neutrino mixing with muon neutrino obtained in this work (dots) from simulations of the HNL signal at PS191 experiment and bounds published by PS191 collaboration. They are obtained for particular processes indicated on the plot. There are also constraints from other experiments (E949\cite{E949:2014gsn}, T2K\cite{T2K:2019jwa} on the same mixing.}
    \label{fig:mumu}
\end{figure}
\begin{figure}[!htb]
    \centering
    \includegraphics[width=0.8\textwidth]{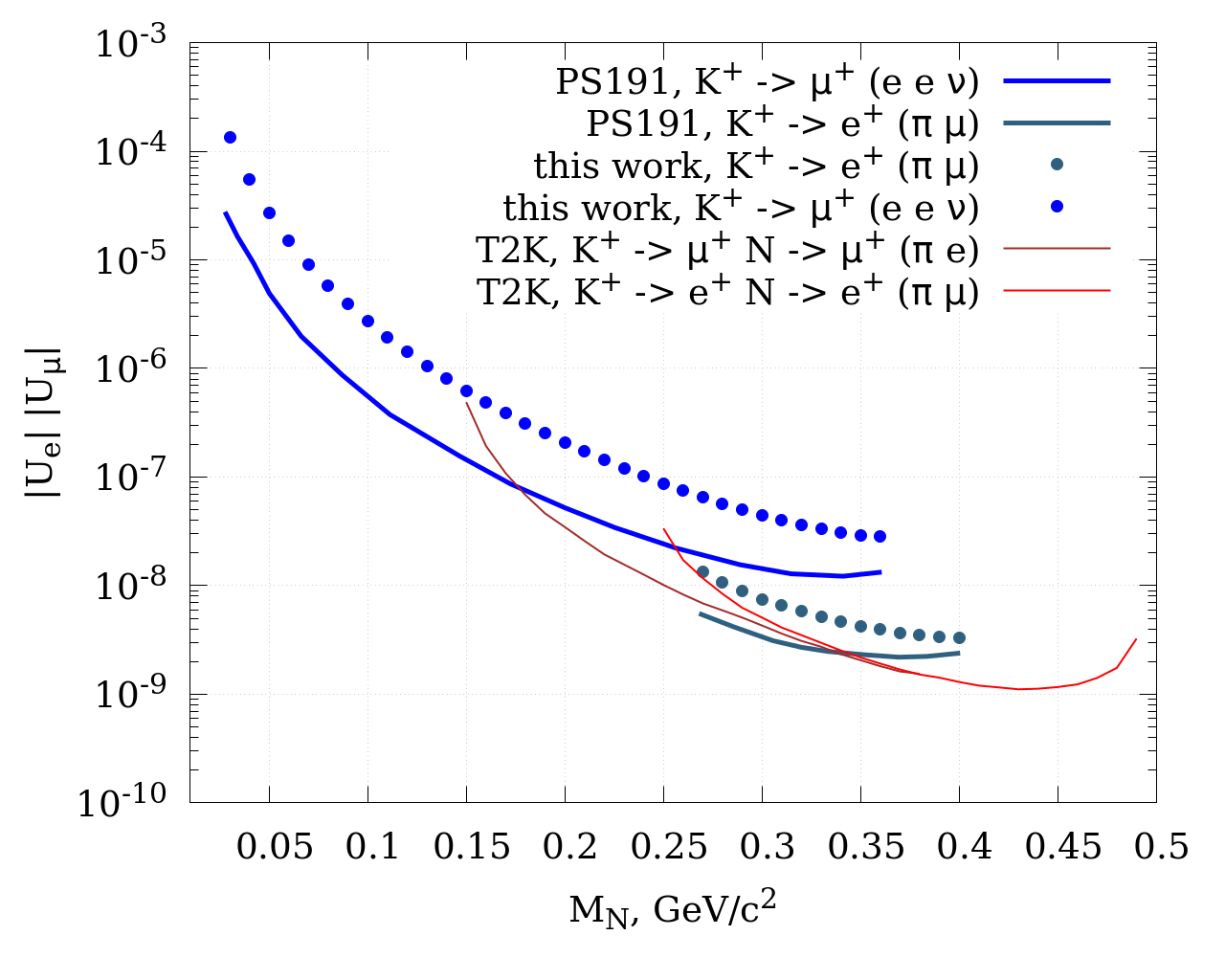}
    \caption{Limits on the product of sterile neutrino mixings with electron and muon neutrinos obtained in this work (dots) from simulations of the HNL signal at PS191 experiment and bounds published by PS191 collaboration. They are obtained for particular processes indicated on the plot. There are also constraints from T2K\cite{T2K:2019jwa} on the same combination of mixing parameters.}
    \label{fig:emu}
\end{figure}
\ref{fig:mumu} and \ref{fig:emu} respectively. Our estimates are outlined with dots, while the published in PS191 papers\,\cite{Bernardi:1985ny,Bernardi:1987ek} bounds and constraints, obtained by the later experiments, are depicted by lines. We sum signals for the both ``on''- and ``off''-modes. Note that we present results for the case of Dirac HNL, and therefore we have corrected the T2K limits by a factor of $\sqrt{2}$ (as they assumed Majorana case).

The PS191 collaborations presented the results for the entire statistics but separately for different pairs, each consisting of particular production and particular decay channels, which we indicated on the plots. In our estimates, we follow the same logic. All the production channels, i.e. $K^+\to e^+N$ and $K^+\to\mu^+ N$ are initiated by the charged current interactions. Their production rates are proportional to $|U_e|^2$ and $|U_\mu|^2$ because of the kaon branching ratios to HNL\,\eqref{Br-2}.  Then, two-body HNL decays into pion and charged leptons, i.e. $N\to\pi^+ e^-$ and $N\to \pi^+\mu^-$ are also initiated by the charged current interactions and the corresponding HNL decay rates in the detector are proportional to  $|U_e|^2$ and $|U_\mu|^2$, as follows from eqs.\,\eqref{2-body-e}, \eqref{2-body-mu}. One can check, that for the interesting values mixing the HNL decay length well exceeds the distance from the target to the detector. Then the signal rates are proportional to $|U_\alpha|^2\times|U_\beta|^2$, where $\alpha$ and $\beta$ stands for electron or muon depending on the type of the charged leptons which attend the production and decay processes. They can be the same, and one can put limits on the product of the same mixing as in Figs.\,\ref{fig:ee} and \ref{fig:mumu}, or they can be different, which yields the limits on the products of different mixing as in Fig.\,\ref{fig:emu}. 

Recall that calculating the HNL decay rates into the three-body final states, i.e. $e^+e^-\nu$, $\e^-\mu^+\nu$ and $\e^+\mu^-\nu$ the PS191 collaboration accounted only contributions of the charged currents, while the neutral currents also contribute to the first processes as we discussed above. The neutral current contribution reduces the decay rate of $N\to \nu_e e^+ e^-$ by a factor $\eta_{NC}$\,\eqref{corr-CC-NC}. To correct for this factor, we rescaled the original PS191 limits indicated as ``PS191, $K^+\to e^+(e\,e\,\nu)$'' in Fig.\,\ref{fig:ee} by a factor of $1/\sqrt{\eta_{NC}}\approx 1.3$, making them a bit weaker. The same correction we make for the limit indicated as ``PS191, $K^+\to \mu^+(e\,e\,\nu)$'' in Fig.\,\ref{fig:emu}. Let us notice, that initiated by the neutral currents decay $N\to e^+ e^- \nu_\mu$ also contribute to the same final state. The contribution is proportional to $|U_\mu|^2$, while the decay rate of $N\to \nu_e e^+ e^-$ is proportional to $|U_e|^2$. At the same value of mixings, the former, see eq.\,\eqref{3-body-muee}, is lower than the latter, see eq.\,\eqref{3-body-eee}, by a numerical factor of about 5. Hence the rescaled by the factor  $1/\sqrt{\eta_{NC}}\approx 1.3$  limit ``PS191, $K^+\to \mu^+(e\,e\,\nu)$'' is reliable only in the models where $|U_\mu|^2<5|U_e|^2$. In models with opposite hierarchy, this limit rather refers to the pure mixing with muon neutrino, $|U_\mu|^2$. Indeed, in the absence of any mixing with electron neutrino (in this way all the bounds on $|U_\mu|^2$ in Fig.\,\ref{fig:mumu} are derived) this channel implies the limit on $|U_\mu|^2$ which can be obtained by rescaling of the limit  ``PS191, $K^+\to \mu^+(e\,e\,\nu)$'' by the numerical factor 2.8. One can check it does not improve the PS191 constraints in Fig.\,\ref{fig:mumu}. Finally, one can also study the same final state but another production, via $K^+\to e^+ N$, which would allow placing a limit on the same product $|U_e U_\mu|$ if $|U_\mu|^2>5|U_e|^2$. We expect it to be weaker than those from the two-body HNL decays already presented in Fig.\,\ref{fig:emu}.    

However, all these corrections do not help to understand why our  estimates, indicated by dots, always show noticeably weaker bounds on $|U_e|^2$ and $|U_\mu|^2$ than the original (and corrected as explained above) limits by PS191. It seems in our simulations we either perform accurate calculations or make approximations which would only increase the PS191 sensitivity to HNL. In other words, we conclude, that the true bounds, which PS191 could obtain, should be above our estimates. Where exactly, we cannot say from our simplified analysis. Nevertheless, it is not important now, since all our limits are placed above the bounds of the later experiments. Since the true PS191 limits should be above ours, their exact form does not matter, since they have been superseded by the later experiments. Limits on $|U_eU_\mu|$ are not placed by most nowadays experiments as rather model-dependent and overall limited by combined analysis of limits on $|U_e|^2$ and $|U_\mu|^2$.

{\bf 5.} To conclude, we simulated the HNL production and decay within the setup of the PS191 experiment and find that their published limits are most probably too optimistic. It seems that the best limits they could obtain are now surpassed by later experiments, and hence there is no need to study  further the PS191 performance.

\vskip 0.3cm
We thank M.\,Hostert and R.\,Shrock for the valuable correspondence. The work is partly supported by the Russian Science Foundation  RSF grant 17-12-01547.

\bibliographystyle{utphys}
\bibliography{refs}
\end{document}